\documentclass[
prl,twocolumn
]{revtex4}
\usepackage[all]{xy}
\usepackage{latexsym}
\usepackage{amsmath}
\usepackage{amsfonts,amssymb}

\newcommand{\na}{\nabla}
\newcommand{\ce}{{\cal{E}}}
\newcommand{\fh}{{\mathfrak{H}}}
\newcommand{\beq}{\begin{equation}}
\newcommand{\eeq}{\end{equation}}
\newcommand{\bit}{\begin{itemize}}
\newcommand{\eit}{\end{itemize}}
\newcommand{\ben}{\begin{enumerate}}
\newcommand{\een}{\end{enumerate}}
\newcommand{\la}{\langle}
\newcommand{\ra}{\rangle}

\newcommand{\bs}{\boldsymbol}
\newcommand{\f}{\frac}

\newcommand{\mb}{\mbox}

\begin{document}
\title{Derived Category Structure and Non-linear Potential of Gauged S-Duality}

\author{Eiji Konishi}

\address{Faculty of Science, Kyoto University, Kyoto 606-8502, Japan}
\email{konishi.eiji@s04.mbox.media.kyoto-u.ac.jp}
\date{\today}
\begin{abstract}Based on the modeling of type IIB string theory vacua using gauged S-duality and the Chan-Paton symmetries by introducing an infinite number of open string charges (affinization), we derive the derived category formulation of the quantum mechanical world including gravity. This leads to the concept of a non-linear potential of gauged and affinized S-duality which specifies the morphism structure of this derived category.

\end{abstract}
\maketitle
%\section{Introduction}
%
{\it $\S .1$ Introduction.}
 Since quantum mechanics was first conceived, the quantum mechanical world (also referred to in the following just as `the world') has needed a new language to describe it and to resolve its measurement problem. In Penrose's thesis\cite{Penrose}, quantum mechanical world branching, that is, the collapse of the superposition of wave functions, in a measurement process is a phenomenon induced by the effects of quantum gravity on the cosmic time. The traditional way to quantize gravity has relied on the classical variables of a space-time metric tensor and matter fields.\cite{WDW1,WDW2,WDW3,WDW4,WDW5} One expects that in a genuinely quantized theory of gravity these classical ingredients are substantially removed and replaced by a discrete kind of geometry. In this letter, based on a string theory view point, we propose a novel way to describe the quantum mechanical world, including gravity, by using the notion of a derived category. This description is the logical consequence of previous work on modeling type IIB string theory vacua by gauged and affinized S-duality.\cite{Konishi1} As our scheme, we first formulate the world as the derived category constructed from a given wave function with the trivial morphism structure; second, we reformulate it by introducing a non-linear potential instead of the wave function. This potential specifies the morphisms of the category as the transformation operators compatible to it (namely, where it is a vector). The cohomological information of the objects of the category results from their interrelationships.

We start with some preliminary explanations of the theory of gauged S-duality.\cite{K,Konishi1}

The Neveu-Schwarz-Ramond (NSR) model of type IIB string theory contains the massless and bosonic excitations of the axion $\hat{\chi}$, dilaton $\hat{\phi}$, graviton $\hat{g}_{MN}$, 2-form Neveu-Schwarz-Neveu-Schwarz (NS-NS) and Ramond-Ramond (R-R) potentials ($\hat{B}^{(i)}_{MN}$ for $i=1,2$, respectively) and the R-R 4-form potential with its self-dual field strength. A hat indicates that a field is ten-dimensional. The effective action in the Einstein frame is\cite{hull,jhs}
 \begin{eqnarray}
&&
S=\frac{1}{2\kappa^2}\int d^{10}x\sqrt{-\hat{g}}
\times\nonumber\\&&
\biggl[\hat{R}_{\hat{g}}+\frac{1}{4}{\rm{Tr}}(\partial_M\hat{{\cal{M}}}\partial^M\hat{{\cal{M}}}^{-1})-\frac{1}{12}{\hat{{\boldsymbol{H}}}}_{MNP}^T\hat{{\cal{M}}}{\hat{{\boldsymbol{H}}}}^{MNP}\biggr]\;,\label{eq:NSR}
\end{eqnarray}
where the $SL(2,{\boldsymbol{R}})/SO(2)$ moduli matrix $\hat{{\cal{M}}}$ of the axion-dilaton and the vector of $H$-fields are
\begin{equation}
\hat{{\cal{M}}}=\left(\begin{array}{cc}\hat{\chi}^2e^{\hat{\phi}}+e^{-\hat{\phi}}&\hat{\chi}e^{\hat{\phi}}\\ \hat{\chi} e^{\hat{\phi}}&e^{\hat{\phi}}\end{array}\right)\;,\ \ {{\hat{{\boldsymbol{H}}}}}_{MNP}=\left(\begin{array}{cc}\hat{H}^{(1)}\\ \hat{H}^{(2)}\end{array}\right)_{MNP}\;,
\end{equation}
and $\hat{H}^{(i)}=d\hat{B}^{(i)}$.
We exclude the R-R 4-form potential from consideration.
The action in Eq.(\ref{eq:NSR}) is manifestly invariant under the S-duality transformations
\begin{eqnarray}\hat{{\cal{M}}}\to\Lambda\hat{{\cal{M}}}\Lambda^T\;,\ \ {\hat{{\boldsymbol{H}}}}\to (\Lambda^T)^{-1}{\hat{{\boldsymbol{H}}}}\;,\ \ 
\hat{g}_{MN}\to \hat{g}_{MN}\;,\label{eq:Sgroup}
\end{eqnarray}where $\Lambda\in SL(2,{\boldsymbol{R}})_S$. 

According to \cite{Konishi1}, we gauge and quantize the S-duality group of Eq.(\ref{eq:Sgroup}).
We regard the pair of the axion and dilaton and that of F- and D-strings as the gauge bosons of gauged S-duality. Since the axion and dilaton parameterize the coset $SL(2,{\bs{R}})/SO(2)\simeq{\fh}$, the corresponding gauge potentials, as the connections on the fiber bundle, satisfy\begin{equation}a_n\in d{\fh}\;,\end{equation}
for the tangent space of the Poincar\'e upper half plane $d{\fh}$ and index $n$ of the base space coordinates $s_n$. 
However, we assume that the gauge group is originally $SL(2,{\bs{R}})_S$ and the Lie algebraic constraints on the field operators reduce it to the coset $SL(2,{\bs{R}})_S/SO(2)$. Then, the number of the generators of the gauge symmetry is still three. 
In type IIB string theory, the S-duality gauge group of Eq.(\ref{eq:Sgroup}) is quantized from $SL(2,{\bs{R}})_S$ to $SL(2,{\bs{Z}})_S$ by imposing Dirac's charge quantization condition on the charges. To achive it, we restrict the coordinates $s_n$ to be discrete, so that the infinite dimensional unitary representations of the transformation operators belong to the representation of $SL(2,{\bs{Z}})_S$. When we refer to the coordinates $s_n$ on the base space in type IIB string theory (not in type IIB supergravity), we regard them as discrete variables.

We construct a model of type IIB string theory vacua by introducing an infinite number of open string charges of Chan-Paton symmetries on the systems of fundamental (F-) strings and multiple Dirichlet (D-) strings with arbitrary multiplicities besides the infinitesimal generators of S-duality\footnote{In the following, we refer the representations of the charges, which are the generators of the symmetry group, merely as the `generators'.}
\begin{equation}
sl(2,{\bs{R}})_{{{S}}}=\la\Sigma^i(0);\ i=0,1,2\ra_{{\bs{R}}}\;.
\end{equation} 
and affinizing to the $\hat{\mathfrak{g}}=\widehat{sl(2,{\boldsymbol{R}})_S}$ algebra (i.e., incorporating the world sheet degrees of freedom (d.o.f.) of a perturbative string theory into an affine Lie algebra based on $sl(2,{\boldsymbol{R}})_{{{S}}}$ algebra)
\begin{subequations}
\begin{align}
&[\Sigma^1(l_1),\Sigma^2(l_2)]=[\Sigma^1,\Sigma^2](l_1+l_2)+l_1\delta_{l_1,-l_2}%
\times\nonumber\\&
(\Sigma^1,\Sigma^2)z\;,\label{eq:affine}\\
&[z,\hat{\mathfrak{g}}]=0\;,\label{eq:affine2}\\
&[\partial,\Sigma(l)]=l\Sigma(l)\;,\label{eq:affine3}
\end{align}
\end{subequations}for $l\in{\boldsymbol{Z}}$.
By gauging these symmetries, we consider the generalized Yang-Mills theory with the gauge affine Lie algebra $\hat{{\mathfrak{g}}}$ of the gauge field and Faddeev-Popov ghost and anti-ghost fields on the $\hat{\mathfrak{g}}$ fiber bundle over an infinite dimensional base space. The coordinates of the base space are an infinite number of time variables $(s_n)_n$ with $n\in{\boldsymbol{Z}}$.\cite{Konishi1} 
We denote the coupling constant of the $\hat{\mathfrak{g}}$ Yang-Mills theory by $\hat{g}$ and identify it with the modulus parameter of F-string world sheets in the weak string coupling region (if necessary by taking S-duality).\cite{Konishi1} The gauge bosons of this theory represent the Chan-Paton gauge bosons. After gauge fixing, this theory has two super gauge symmetries: the Bechi, Rouet, Stora and Tyutin (BRST) and anti-BRST symmetries.\cite{BRST1,BRST2,BRST3} Then, the Kugo-Ojima physical state condition on the wave function\cite{KO}
\begin{equation}
Q\Psi=0\;,\ \ Q^2=0\;,\label{eq:brst}
\end{equation}
for the sum of the BRST and anti-BRST charges $Q$ holds.
The solution of Eq.(\ref{eq:brst}) is a Kac-Peterson form of a theta function.\cite{Konishi1} The modular invariance of this theta function about the coupling constant $\hat{g}$ with modular group $\Gamma$ reflects the T-duality invariance of the wave function.
Here, $\Psi$ is a function of the time variables $(s_n)_n$ and the coupling constant $\hat{g}$.
We can interpret this wave function $\Psi$ as the wave function of the Universe. Actually, it contains all excitations of F- and D- strings except for gravitons and the excitations of R-R 4-form potential. The space-time geometry is embedded into fiber space as the probe of spatial d.o.f. by F- and D- strings.\footnote{This process is done according to the manner by Gliozzi, Scherk and Olive\cite{Konishi1,Modular1,Modular2} and compatible with the NSR algebra that\cite{NSR} indicates the Lorentz signature of the ten-dimensional space-time metric. So, we assume this signature of the space-time metric.} So, due to the Eguchi-Kawai large $N$ reduction applied to the infinite number of gauge generators of $\hat{\mathfrak{g}}$, the dynamical d.o.f. on the fiber space are completely reduced. That is, the d.o.f. of type IIB string theory appearing in the field and metric configurations are reduced to an integrable hierarchy of the time variables $(s_n)_n$ on the Riemann surface of the coupling constant $\hat{g}$.

 Our model has two distinct structures. First structure is the gauge field theory on the moduli space of vacua. Each vacuum is specified by the Kugo-Ojima physical state condition and is the stable field configuration when we regard the BRST charge as a differential. As will be explained, this gauge field theory is just an infinitesimal local description of the moduli space of vacua in type IIB string theory. Non-perturbative effects are not yet described and non-perturbative field configurations are fixed and not dynamical. So, the contents of this stage of the modeling are not so different from those of a perturbative string theory. The non-perturbative description or dynamics of type IIB string theory, i.e., transition between the stable configurations, is achieved by introducing another non-linear potential, as the second structure of the model, which can describe the moduli space of vacua globally. Then, we can describe non-perturbative effects, such as an infinite many body effect and the dynamics of D-branes. The way to introduce the second gauge potential is based on the derived category structure of the state spaces generated by a fixed vacuum. This derived category structure bases on Eq.(\ref{eq:tau}) and results from the perturbative string symmetry, T-duality. Due to these structures, the theory studied in this letter is, on the whole, equivalent to the standard non-perturbative formulation of type IIB string theory\cite{IKKT} including the issue of the unitarity.

We define the renormalization of the wave functions by the methods of Whitham deformation $R$\cite{Wh} of the variables $p$ that are canonically conjugate to the variables of the representation space of the field operators: \begin{equation}p\to P(S)\;,\ \ s=\epsilon^{-1}I(S)\;,\end{equation} for the slow variables $S$, the scale constant $\epsilon$ and their multi-phase function $I(S)$, so that $R$ satisfies the conservation law\begin{equation} QR\Psi=0\;.\label{eq:Ren}
\end{equation} 
 We define the matter field $\psi_\Lambda$ with order parameters $\Lambda$ and renormalization $R$ by
\begin{equation}
\psi_\Lambda=R\Psi_{\hat{v}}\;,\ \ \Psi_{\hat{v}}=\varrho_1(\hat{v})\Psi\;,\label{eq:system}
\end{equation}
for the $U(\hat{\mathfrak{g}})$ representation $\varrho_1$ restricted to an element ${v}$ of the $U(\hat{\mathfrak{g}})$-weight module $V$. {The highest weight vector is implicitly contained in the wave function $\Psi$.\cite{Konishi1}} We denote the subspace of the $U(\hat{\mathfrak{g}})$-weight module $V$ by $V_s$, which consists of the elements ${v}$ in Eq.(\ref{eq:system}).

The cosmic time $\tau$ is the affine parameter assigned on spatial hypersurfaces sliced from space-time. Our definition of the increment of the cosmic time is done by a clock of the string excitations.
{{We define the increment of the {{cosmic time}} $\delta{\tau}(s)$, to describe the change of the system only, in units of the Planck time as the operator, whose expectation value in the system is for the expectation values of the momenta ${p}^0$\cite{Konishi1}
\begin{equation}\la\delta{\tau}(s)\ra=\biggl[\frac{ k\la Q\ra}{\la \sqrt{(\Omega_0-\Omega)({p}^0)}\ra}\biggr](s)\;,\label{eq:tau}\end{equation}where the numerator of Eq.(\ref{eq:tau}), $k{Q}(s)$, is proportional to the Hamiltonian of the system ${Q}(s)$ 
 as the frequency times the number of the elements of the system (when we consider the free part of it),
and the denominator of Eq.(\ref{eq:tau}) is defined by the square root of the shifted minus Casimir operator $\Omega$ of the representation of the affine Lie algebra $\hat{{\mathfrak{g}}}$ in the field operators,\cite{Casimir,VGKAC}
where we invoke the identity for an arbitrary functional $f$ of ${\Omega}$
\begin{eqnarray}
\la w|f({\Omega})\delta \tau|w\ra&&=\sum_{w^\prime}\la w|f({\Omega})|w^\prime\ra \la w^\prime|\delta \tau|w\ra\nonumber\\
&&=\la w|f({\Omega})|w\ra\la w|\delta \tau|w\ra\;.
\end{eqnarray}
In Eq.(\ref{eq:tau}) $\Omega_0$ is the maximum value of the Casimir operator.

The coordinates $s$ have no relevance to the history of the cosmic time $\tau$ but only its increment $\delta\tau(s)$.
%\begin{equation}\delta\tau(s)=\int^s ds^\prime\delta\hat{\tau}(s^\prime)\;.\label{eq:time}\end{equation}
 The solutions $\Psi$ of Eq.(\ref{eq:brst}) do not have the variable of the history of the cosmic time $\tau$, so we add it to $\Psi$.

These definitions are equivalent to the equation
\begin{equation}\sqrt{ (\Omega_0-\Omega)({p}^0)}\delta {\tau}(s)= k{Q}(s)\;.\label{eq:grav1}
\end{equation}
We notice that the Casimir part of Eq.(\ref{eq:tau}) has dimensions of time, since $\hbar$ times the square root of the shifted minus Casimir of the representation in the field operators (that is, the sum of $\hbar$ times the time frequencies of string excitations $\omega_{p^0}$ in the field operators) has dimensions of energy. The string excitations and the time periods corresponding to these frequencies $\omega_{p^0}$ are recognized as the clock and the Casimir part of  Eq.(\ref{eq:tau}) respectively.
Thus, the free part of Eq. (\ref{eq:tau}) is proportional to the quotient of the expectation value of the Hamiltonian divided by the time frequency of the string excitation $\omega_{p^0}$, that is, the expectation value of the number of the elements of the system.

 With respect to the free part of the Hamiltonian, the grounds for the definition in Eq.(\ref{eq:tau}) is as follows. As explained above, the free part of Eq.(\ref{eq:tau}) is proportional to (the expectation value of) the number of elements of the system $\la n\ra$. When we fix the expectation values of momenta $p^0$, the statistical properties of $\la n\ra(s)$ around the coordinates $s$ are those of the eigenstates of the Hamiltonian $|n\ra$ (if we restrict the Hamiltonian to its free part) which are labeled by the numbers of elements $n$. The cosmic time increment needs to count all of the non-unitary processes. The non-unitary processes induced by the Hamiltonian are classified by the transitions between the states labeled by the numbers $\la n\ra$ via their superpositions. Thus, the variance of the number $\la n\ra(s)$ around the coordinates $s$ captures the changes in the non-unitary processes induced by the Hamiltonian. Consequently, $\la n\ra$ contains the statistical properties (i.e., mean, variance and distribution function) of $\delta\tau(s)$ around the coordinates $s$. Here, we have discussed Eq.(\ref{eq:tau}) using coarse-grained values of some quantities, so Eq.(\ref{eq:tau}) may not be the exact form of the cosmic time increment. On this issue, further refinement may be needed, but is beyond the scope of the present investigations.

The cosmic time development of the wave function is governed by
\begin{equation}
i\hbar\frac{\delta \psi_\Lambda}{\delta \tau_R}={\cal{H}}\psi_\Lambda\;,\ \ {\cal{H}}=QR|_{V_s}\;,\label{eq:Sch}
\end{equation}
for the effectively scaled increment of the cosmic time $\delta\tau_R=\delta\tau|_{RV_s}$. We note that $\delta/\delta \tau_R=\delta/\delta(\delta\tau_R)$.

The increment of the cosmic time is an operator, thus it makes a rigorous sense only for its eigen wave functions. In this paper, we consider the time developments of its eigen wave functions only, otherwise we take its expectation value in the system.

The functional form of the increment of the cosmic time $\delta\tau_R(s)$ is already determined by Eq.(\ref{eq:tau}). Eqs. (\ref{eq:Sch}) specify the functional variation $\delta/\delta \tau_R$. The time coordinates $s$ are infinitely many, matching the number of degrees of freedom of the model. This variation $\delta\psi_\Lambda/\delta \tau_R$ is between functions of the coordinates $s$. Without Eqs.(\ref{eq:Sch}), when $\tau_R$ changes its value, we do not know which of the coordinates $s$ has changed to cause this shift. Eq.(\ref{eq:Sch}) specifies it. The cosmic time is a degree of freedom among the infinite number of coordinates. If we fix the cosmic time, the wave function is a function of the remaining degrees of freedom. The variation of the wave function with respect to the cosmic time is determined by Eqs. (\ref{eq:Sch}).

The cosmic time has its sense only as its change in Eq.(\ref{eq:Sch}) of the corresponding state space $V_s$.
Then, due to Eq.(\ref{eq:brst}) the bare wave function does not depend on the cosmic time
\begin{equation}
\delta_{\tau}\Psi=0\;.\label{eq:est1}
\end{equation}
In the substate space $V_s$, the description of the cosmic time is incomplete and stochastic.
The solution of Eq.(\ref{eq:Sch}), when the cosmic time dependence is averaged over the coordinates on the moduli space of vacua, is
\begin{equation}
\langle\psi_{\Lambda}(\tau_R)\rangle\approx\exp\biggl(-\frac{i\tau_R}{\hbar} {\cal{H}}-\frac{\sigma_R\tau_R}{2\hbar^2}{\cal{H}}^2\biggr)\langle\psi_{{\Lambda}}(0)\rangle\;,\label{eq:est2}\end{equation}
where the average is defined by the recursion equation
\begin{subequations}
\begin{align}
\langle\psi_\Lambda(\mu_R)\rangle&=\int{\cal{D}}\tau^\prime_R(s)\exp\biggl(-\frac{i\delta\tau_R^\prime(s)}{\hbar}{\cal{H}}\biggr)\langle \psi_{\Lambda}(0)\rangle\\
&\approx\int d\tau_R^\prime \exp\biggl(-\frac{i\delta\tau_R^\prime}{\hbar}{\cal{H}}\biggr)f(\delta\tau_R^\prime)\langle \psi_{\Lambda}(0)\rangle\;.
\end{align}
\end{subequations}
Here, we rewrite the functional integral with respect to $\delta\tau_R^\prime (s)$ as an average over a normal stochastic variable $\delta\tau_R$ with variance $\sigma_R$, mean $\mu_R$ and distribution function $f(\delta\tau_R^\prime)$.
We note that, in general, ${\cal{H}}^2$ is not identically zero due to its restriction.
In Eq.(\ref{eq:est2}), the first and second factors are the unitary and non-unitary processes, respectively. This is a string theory realization of Penrose's idea.

%\section{Derived Category Structure Using Wave Functions}
%
{\it {$\S .2$ Derived Category Structure Using Wave Functions.}}
In our modeling, due to Eqs.(\ref{eq:est1}) and (\ref{eq:est2}), the perfect description of the Universe is independent of changes in the cosmic time, and non-trivial cosmic time processes can be applied only to closed systems with imperfect, partial descriptions and a non-zero retention time of the superposition of the wave functions.
{Due to Eq.(\ref{eq:est2}), the retention time of the superposition of the wave functions tends to zero for the macroscopic objects.} Systems which lose the retention time of the superposition of the wave functions
 have a classical cosmic time evolution and are essentially removable objects, whereas systems with a non-zero 
 retention time of the superposition of the wave functions genuinely constitute a quantum mechanical world with common cosmic time processes such as quantum mechanical branching. That is, for the system with the non-zero retention time of the superposition of the wave functions, the variance of the increment of the cosmic time induces the non-unitary time development of a system. By Eq.(\ref{eq:system}) a system is a state space $RV_s$ of a $U(\hat{\mathfrak{g}})$-module $V_s$ with a certain renormalization $R$ and its time development is mapped to the projective resolution of the diagram of\begin{equation}\xymatrix{0& RV\ar[l]_Q}\end{equation} as the $Q$-complex \begin{equation}\xymatrix{0& RV\ar[l]_Q& P_1RV\ar[l]_Q&\cdots\ar[l]_Q}\label{eq:exact}\end{equation}where in the $n+2$-th element of Eq.(\ref{eq:exact}) we restrict both of $QR$ and $\Psi$ to the same state space with the $n$-th cosmic time value counted by the events of non-unitary processes.

 Since in our context, the Kugo-Ojima physical state condition means that the wave function is an eigenfunction of the Hamiltonian with zero eigenvalue, the non-unitary time process, that is, the collapse of a superposition of wave functions changes the eigenvalues of the Hamiltonian. Namely, for the eigenvalue $\lambda$ of the Hamiltonian
 \begin{equation}
 {\mb{ker}}Q=RV_{\lambda=0}\;,\ \ {\mb{im}}Q=RV_{\lambda\neq0}\;,
 \end{equation}
 holds. Then, the kernel of $Q$ does not match the full state space $V$ and the cohomology of $Q$ is non-trivial, that is, not the full state space. We note that, generally, superpositions of the wave functions are generated by unitary time processes in the larger system. Thus, a non-unitary process may occur at any cosmic time.

 By the $Q$-cohomology content in Eq.(\ref{eq:est2}) only, each system is specified and the cohomologically non-trivial content is the non-unitary second factor in Eq.(\ref{eq:est2}) which survives modulo Im$Q$. For a macroscopic physical object, we can interpret this as a collection of microscopic quantum states with non-trivial effects of time variances or as a large-scale macroscopic quantum state with trivial effect of time variance. 
These interpretations need to be unified.
These observations lead us to the derived category description of the quantum mechanical world under the moduli of quasi-isomorphism equivalences of the BRST complexes. We denote by $D^b(C)$ this derived category of the BRST complexes of the base abelian category $C$.
   Due to Eq.(\ref{eq:Ren}), the quasi-isomorphisms which commute with the cosmic time development by the $Q$ operation are given by renormalizations. 
Here, the derived category $D^b(C)$ of a base abelian category $C$ is defined by restriction of the homotopy category $K(C)$ on a closed system of the products of quasi-isomorphisms in $K(C)$.\cite{GM}
 The objects of our base abelian category $C$ are the spaces of states created from a given wave function $\Psi$ by the actions $\varrho(V)$, i.e., the $U(\hat{\mathfrak{g}})$-weight modules. The morphisms are the transformations compatible with the differential $Q$ or the covariant derivative $\nabla$ (namely, where $Q$ or $\nabla$ is a vector) respectively. (We denote the latter base abelian category by $C^\nabla$.)
When we consider the theory of gauged S-duality using the linear wave function $\Psi$ only, the morphisms of base abelian category $C^Q$ are defined only by the homomorphisms, denoted by $k$, compatible with the differential $Q$. Then, the morphisms, denoted by $h$, between two complexes are defined by $h=kQ+Qk$.

%\section{Non-perturbative Description Using the Non-linear Potential}
%
{\it $\S .3$ Non-perturbative Description Using the Non-linear Potential.}
Based on this derived category structure $D^b(C)$ of the quantum mechanical world description, we generalize the results in the Introduction 
by a substantially different method. We introduce a single master equation as the generalization of the Kugo-Ojima physical state condition for a non-linear potential, denoted by $\aleph$ (standing for the symbol `$A_\mu$' of a gauge potential), which describes the non-peturbative effect or dynamics alluded in the Introduction,
 according to the following three guiding principles. As the concrete form of the equation, we adopt a single vanishing curvature.
 
\renewcommand{\theenumi}{\roman{enumi}}
\renewcommand{\labelenumi}{\theenumi)}

\begin{enumerate}
\item
{{The local principle.}} In our modeling, it is gauged S-duality.
\item The generalized gauge invariance using the non-linear potential.
%The homotopical covariance in the derived category $D(C)$ of the $Q$-complexes.
\item  The equation vanishes under the action of the covariant derivative, due to the generalized gauge invariance (we note that the covariant derivative is the generalization of the BRST charge $Q$).
\end{enumerate}

As the result, the generalized Kugo-Ojima physical state condition is regarding operator valued $\hat{\mathfrak{g}}$-connection $\aleph$ on the fiber space:\footnote{In the following, we denote the BRST charge and the anti-BRST charge by $Q$ and $\bar{Q}$ respectively. So, the notation $Q$ used here is different from that in Eq.(\ref{eq:Q}).}
\begin{equation}\Omega-\frac{1}{2}[[\bar{\na},\Omega],\na]=0\;,\label{eq:master}\end{equation}
where $\Omega$ is the curvature form
\begin{equation}
\Omega=[\na,\na]\;,
\end{equation} and we introduce $\bar{\aleph}$ as the dual vector field, whose coefficients $\bar{\aleph}_i$, indexed by the differential basis $\partial_i$ on the tangent bundle for the dual coordinates ${\cal{Y}}^i$ of $Q_i=Q|_{dy^i}$, are the operators made from the dual basis of $\hat{{\mathfrak{g}}}$. These coefficients are canonically conjugate to the one of $\aleph$, indexed by the 1-form basis $d{\cal{Y}}^i$ on the cotangent bundle, as \begin{equation}[{\aleph}_i,\bar{\aleph}_j]=\delta_{ij}\;,\label{eq:can}\end{equation} and $\nabla$ and ${\bar{\nabla}}$ are the covariant derivatives on the categorical $Q$-complexes
\begin{equation}\nabla{\cal{O}}=\delta{\cal{O}}+[\aleph,{\cal{O}}]\;,\ \ \bar{\nabla}{\cal{O}}=\bar{\delta}{\cal{O}}+[\bar{\aleph},{\cal{O}}]\;,\end{equation}for an arbitrary operator valued form ${\cal{O}}$ and $[,]$ is the ${\boldsymbol{Z}}$-graded 
commutator for an arbitrary pair of a $d_a$-form ${\cal{O}}_a$ and $d_b$-form ${\cal{O}}_b$:\footnote{The action of the differential basis on a commutator of forms satisfies a Leibniz rule similar to the one for $Q$ (see Eq.(\ref{eq:Leibniz})). It is consistent due to $d{\cal{Y}}^i\wedge d{\cal{Y}}^i=0$.}
\begin{equation}[{\cal{O}}_a,{\cal{O}}_b]={\cal{O}}_a\wedge {\cal{O}}_b-(-)^{d_ad_b}{\cal{O}}_b\wedge {\cal{O}}_a\;,\end{equation}
which satisfies the super Jacobi identity
\begin{eqnarray}
&&
(-)^{d_ad_c}[{\cal{O}}_a,[{\cal{O}}_b,{\cal{O}}_c]]+(-)^{d_bd_c}[{\cal{O}}_c,[{\cal{O}}_a,{\cal{O}}_b]]+
\nonumber\\&&
(-)^{d_ad_b}[{\cal{O}}_b,[{\cal{O}}_c,{\cal{O}}_a]]
=0\;.\label{eq:Jacobi}
\end{eqnarray}
Here, we consider the super Lie algebra of $\hat{{\mathfrak{g}}}$, and the degree $d$ of the element ${\cal{O}}$ is its ghost number. We note that in general, ${\cal{O}}_a\wedge {\cal{O}}_b\neq-{\cal{O}}_b\wedge{\cal{O}}_a$ for $1$-forms, since we treat the product of matrices in $\hat{{\mathfrak{g}}}$ and the outer product of forms simultaneously. Since the BRST differential has ghost number one, the ghost number coincides with the degree of the element as a form. The space of the operators $O$ splits into $\oplus_{i\ge0} O^i$ labeled by the ghost number $i$ with $[O^i,O^j]\subset O^{i+j}$. The BRST differential shifts $O^i$ to $O^{i+1}$ and acts on commutators of operators as \begin{equation}\delta[{\cal{O}}_a,{\cal{O}}_b]=[\delta{\cal{O}}_a, {\cal{O}}_b]+(-)^{d_a}[{\cal{O}}_a, \delta{\cal{O}}_b]\;.\label{eq:Leibniz}\end{equation} 

We check the requirements of the three principles in Eq.(\ref{eq:master}).
The first principle requires that infinitesimal deformations $\Psi$ of the parallel section to $\nabla$ obey the linearized equation Eq.(\ref{eq:brst}).
The principle of covariance requires that the non-linear potential $\aleph$ obey the equation written only using $\nabla$ and $\bar{\nabla}$. Eq.(\ref{eq:master}) satisfies these requirements. Finally, to show the third principle on Eq.(\ref{eq:master}), we contract the indices $i$ and $b$ of the Bianchi identity
\begin{equation}
\nabla_i([\nabla_j,\na_k])_{ab}+\nabla_k([\nabla_i,\na_j])_{ab}+\nabla_j([\nabla_k,\na_i])_{ab}=0\;,
\end{equation}for the components of $[\nabla,\nabla]$
 obtained from the super Jacobi identity in Eq.(\ref{eq:Jacobi}). For the components of $\nabla$, we have\begin{eqnarray}
 [\nabla_i,[\nabla_j,\nabla_k]]+[\nabla_k,[\nabla_i,\nabla_j]]+ [\nabla_j,[\nabla_k,\nabla_i]]=0\;,
\end{eqnarray}where we use the fact that $\nabla$ has ghost number $1$. We 
denote the matrix elements (not in the sense of the expectation values) of the operator $[\nabla_i,\na_j]$ by $([{\nabla_i,\na_j}])_{ab}$, and these are defined by \begin{equation}[\nabla_i,\nabla_j]{\cal{O}}^a=([\nabla_i,\na_j])^a_{b}{\cal{O}}^b\;,\end{equation} for an arbitrary operator valued $\hat{{\mathfrak{g}}}$-connection ${\cal{O}}$ on the fiber space. The indices $a$ and $b$ denote the bases of $\hat{{\mathfrak{g}}}$, and the contraction of indices is taken using the metric on the fiber space. By analogy with the $c$-valued curvature tensor, we assume the symmetry of the indices of $[\nabla,\nabla]$
\begin{equation}
([\nabla_i,\na_j])_{ab}=([\nabla_a,\na_b])_{ij}=-([\nabla_j,\na_i])_{ab}=-([\nabla_i,\na_j])_{ba}\;.\label{eq:Riemann}
\end{equation} Then, the Bianchi identity takes the form 
\begin{eqnarray}\nabla^iG_{aijk}
=0\;,\end{eqnarray}
where
\begin{eqnarray}G_{aijk}=&&([\nabla_j,\na_k])_{ai}+\delta_{ij}\sum_i([\nabla_k,\na_i])_{ai}\nonumber\\&&-\delta_{ik}\sum_i([\nabla_j,\na_i])_{ai}\;.\label{eq:G}
\end{eqnarray}
Using the Leibniz rule, the equality \begin{equation}\bar{\delta}{\aleph}+\delta\bar{\aleph}=0\;,\end{equation} and Eq.(\ref{eq:Riemann}) for the interchange of the indices $i$ and $j$ for the action of the component $\bar{\na}_i$ on the component $[\nabla_j,\na_k]$ and using the Leibniz rule and Eq.(\ref{eq:Riemann}) for the interchange of the indices $a$ and $b$ for the action of the dual basis $\bar{S}^i$ of $\bar{\na}=\bar{\na}_i\bar{S}^i$ on the basis $S^a$ and $S^b$ of $[\nabla_j,\nabla_k]=([\nabla_j,\na_k])_{ab}S^aS^b$ indexed by $a$ and $b$ such that $\bar{S}^aS^b=\delta_{ab}$ (the factor $1/2$ in the second term of Eq.(\ref{eq:master}) comes from this action), and keeping in mind the canonical conjugation relations Eq.(\ref{eq:can}),
we find the first term and the sum of the second and third terms of this quantity $G$ to be locally the first and second terms of the l.h.s. of Eq.(\ref{eq:master}) respectively.
Consequently, Eq.(\ref{eq:master}) satisfies the third principle of the generalized gauge invariance.

Based on Eq.(\ref{eq:master}), we define each morphism of the base abelian category $C^{\nabla}$ to be the non-linear\footnote{Of course, this non-linearity is about the element of $U(\hat{{\mathfrak{g}}})$, and each morphism acts on the object linearly.} transformation operator compatible to the non-linear potential $\aleph$ (namely, where the covariant derivative $\nabla$ is an infinite dimensional vector for this transformation just like the situation such that, in the general theory of relativity, the covariant derivative is a vector on the curved space-time, that is, in our case the parallel section $\Psi^\nabla$ of $\nabla$ such that $\nabla\Psi^\nabla=0$). The objects of $C^\nabla$ are redefined to be compatible to the morphisms and do not need a wave function of the Universe, which is an infinitesimal approximation of the parallel section $\Psi^\nabla$ of $\nabla$.
 {{We change the formulation so that the cohomological contents of wave functions result from the morphisms. In this new vision, the role of the given linear potential $\Psi$ in the $Q$-complexes is substantially taken by the non-linear potential $\aleph$ (in the general theory of relativity, they correspond to Newton potential and the space-time metric respectively), and the category $C^Q$ has only cohomologically trivial contents, that is, the vacuum itself or a unitary factor only.}}
The non-linear potential $\aleph$ describes the dynamics of the morphisms of the derived category $D(C)$ which is the morphism structure of base abelian category $C$. This description is global. Consequently, the non-linear potential $\aleph$ can describe the transition between the stable configurations. In contrast, the linear wave function $\Psi$ is an infinitesimal local description of $D(C)$ and $\aleph$, and cannot describe the non-perturbative effect nor dynamics of $D(C)$.


\begin{thebibliography}{99}
\bibitem{Penrose}R. Penrose, Gen. Rel. Grav. {\bf{28}} (1996), 581.
\bibitem{WDW1}J. A. Wheeler, in Relativity, Groups and Topology, eds. C. De Witt and B.S. De Witt, p.315, Gordon and Breach, New York, (1963).
\bibitem{WDW2}J. A. Wheeler, in Batelles Rencontres, eds. C. De Witt and J. A. Wheeler, p. 242, Benjamin, New York, (1968).
\bibitem{WDW3} B.S. De Witt, Phys. Rev. {\bf{160}} (1967), 1113.
\bibitem{WDW4}J. B. Hartle and S. W. Hawking, Phys. Rev. {\bf{D 28}} (1983), 2960.
\bibitem{WDW5}S. W. Hawking, Nucl. Phys. {\bf{239}} (1984), 257.
\bibitem{Konishi1}E. Konishi, preprint, arXiv:1001.3382 [hep-th].
\bibitem{K}E. Konishi,
Prog. Theor. Phys. {\bf{121}} (2009), 1125 arXiv:0902.2565 [hep-th].
\bibitem{hull}C. M. Hull, Phys. Lett. {\bf{B 357}}, 545 (1995), arXiv:hep-th/9506194.
\bibitem{jhs}J. H. Schwarz, Phys. Lett. {\bf{B 360}}, 13 (1995), arXiv:hep-th/9508143.
\bibitem{r1} A. Sen, Int. J. Mod. Phys. {\bf A 9} (1994), 3707 
[arXiv:hep-th/9402002].
\bibitem{r2} A. Giveon, M. Porrati and E. Rabinovici, Phys. Rep. {\bf C 244} (1994), 77 [arXiv:hep-th/9401139].
\bibitem{BRST1}C. Becchi, A. Rouet and R. Stora, Phys. Lett. {\bf{B}} {\bf{52}} (1974), 344; {Ann. of Phys.} {\bf{98}} (1976), 287.
\bibitem{BRST2}I. V. Tyutin, Lebedev Institute preprint (1975, unpublished).
\bibitem{BRST3} N. Nakanishi and I. Ojima, 
{Z. Phys.} {\bf{C 6}} (1980), 155.
\bibitem{KO}T. Kugo and I. Ojima, Phys. Lett. {\bf{B}} {\bf{73}} (1978), 459; {Prog. Theor. Phys. Suppl.} {\bf{66}} (1979), 1.
\bibitem{Modular1}F. Gliozzi, J. Scherk and D. Olive, Nucl. Phys. {\bf{B 122}} (1977), 253.
\bibitem{Modular2}J. Polchinski, Commun. Math. Phys. {\bf{104}} (1986), 37.
\bibitem{NSR}V. G. Kac and I. T. Todorov, Commun. Math. Phys. {\bf{102}} (1985), 337.
 \bibitem{KM}E. Konishi and J. Maharana, Int. J. Mod. Phys. {\bf{A 25}} (2010), 3797 arXiv:0910.4849 [hep-th].
\bibitem{IKKT}N. Ishibashi, H. Kawai, Y. Kitazawa and A. Tsuchiya, Nucl. Phys. {\bf{B 498}} (1997), 467 [arXiv:hep-th/9612115].
\bibitem{Wh}G. B. Whitham, J. Fluid Mech. {\bf{22}} (1965), 273; Proc. R. Soc. London {\bf{A 139}} (1965), 283.
\bibitem{Casimir}H. B. G. Casimir, Proc. Kon. Ned. Acad. Wetenschap, {\bf{34}} (1931), 844.
\bibitem{VGKAC}V. G. Kac, {\it{Infinite-Dimensional Lie Algebras}}, 3rd ed., Cambridge Univ. Press, (1990).
\bibitem{GM}S. I. Gelfand and Y. I. Manin, {\it{Methods of homological algebra}}, 2nd ed., Springer Monographs in Mathematics. Springer-Verlag, Berlin, (2003).
\end{thebibliography}
\end{document}